\pdfoutput=1
\documentclass[11pt]{article}
\usepackage{acl}

\usepackage{times}
\usepackage{latexsym}
\usepackage[T1]{fontenc}
\usepackage[utf8]{inputenc}
\usepackage{microtype}
\usepackage{inconsolata}
\usepackage{tikz}
\usepackage{comment}
\usepackage{ifthen}
\usepackage{tcolorbox}
\tcbuselibrary{breakable}
\tcbuselibrary{theorems}
\tcbuselibrary{skins}
\usepackage{graphicx}
\usepackage{hyperref}
\usepackage[titlenumbered,ruled,vlined]{algorithm2e}
\usepackage[normalem]{ulem}
\usepackage{tabularx}
\usepackage[export]{adjustbox}
\usepackage{amssymb}
\usepackage{amsmath}
\usepackage{enumitem} 
\usepackage{booktabs}
\usepackage{multirow}
\usepackage{xcolor,colortbl}
\usepackage{float}
\usepackage{siunitx}
\usepackage{calrsfs}
\usepackage{caption}
\usepackage{subcaption}

\newcommand{\para}[1]{\paragraph{\textnormal{\textbf{#1.}}}} 

\newboolean{showcomments}
\setboolean{showcomments}{true} %
\ifthenelse{\boolean{showcomments}}
{
	\newcommand{\nb}[3]{
		{\colorbox{#2}{\bfseries\sffamily\scriptsize\textcolor{white}{#1}}}
		{\textcolor{#2}{$\blacktriangleright$\textsf\small{#3}$\blacktriangleleft$}}}
	 
}{
	\newcommand{\nb}[3]{}
}

\newcommand\mybox[2][]{\tikz[overlay]\node[fill=blue!20,inner sep=2pt, anchor=text, rectangle, rounded corners=1mm,#1] {#2};\phantom{#2}}

\newcommand{\gbox}[1]{\mybox[fill=green!20]{#1}}
\newcommand{\bbox}[1]{\mybox[fill=blue!20]{#1}}

\definecolor{DefinitionColor}{RGB}{45, 52, 151}

\newtcbtheorem{definitions}{}%
{enhanced,frame empty,interior empty,colframe=DefinitionColor!50!white,
 coltitle=DefinitionColor!50!black,fonttitle=\bfseries,colbacktitle=DefinitionColor!15!white,
 borderline={0.5mm}{0mm}{DefinitionColor!15!white},
 borderline={0.5mm}{0mm}{DefinitionColor!50!white,dashed},
 attach boxed title to top center={yshift=-2mm},
 boxed title style={boxrule=0.2pt},varwidth boxed title}{defo}

\newcommand{\topdocs}[1]{\theta({#1})_m}

\newcommand{\qrel}[1]{\mathcal{R}(#1)}

\usepackage{tcolorbox}

\usepackage{enumitem} 
\newcommand{\uls}{\begin{itemize}[leftmargin=*,noitemsep,nolistsep]}
\newcommand{\ule}{\end{itemize}}
\newcommand{\ols}{\begin{enumerate}[leftmargin=*,noitemsep,nolistsep]}
\newcommand{\ole}{\end{enumerate}}

\usepackage{colortbl}
\usepackage{xcolor}

\renewcommand{\para}[1]{\paragraph{\textnormal{\textbf{#1}.}}}

\DeclareMathAlphabet{\pazocal}{OMS}{zplm}{m}{n}
\DeclareMathAlphabet{\pazobfcal}{OMS}{cmsy}{b}{n}

\setitemize{noitemsep}
\setlist{topsep=0pt, leftmargin=*}

\let\mathcal\pazocal

\newcommand{\andcg}{\alpha\text{nDCG}}

\author{
  Nilanjan Sinhababu$^{*}$ \\
  Centre for Computational \\
  and Data Sciences \\
  IIT Kharagpur, India \\
  \texttt{\small nilanjansb@kgpian.iitkgp.ac.in}
  \And
  Andrew Parry$^{*}$ \\
  School of Computing Science \\
  University of Glasgow \\
  United Kingdom \\
  \texttt{\small a.parry.1@research.gla.ac.uk}
  \AND
  Debasis Ganguly \\
  School of Computing Science \\
  University of Glasgow \\
  United Kingdom \\
  \texttt{\small Debasis.Ganguly@glasgow.ac.uk}
  \And
  Pabitra Mitra \\
  Department of Computer \\
  Science and Engineering \\
  IIT Kharagpur, India \\
  \texttt{\small pabitra@cse.iitkgp.ac.in}
}

\begin{document}

\title{Modeling Ranking Properties with In-Context Learning}

\maketitle

\def\thefootnote{*}
\footnotetext{These authors contributed equally to this work.}
\def\thefootnote{\arabic{footnote}}

\begin{abstract}
While standard IR models are primarily designed to optimize relevance, real-world search often needs to balance additional objectives such as diversity and fairness. These objectives depend on inter-document interactions and are commonly addressed using post-hoc heuristics or supervised learning methods, which require task-specific training for each ranking scenario and dataset. In this work, we propose an in-context learning (ICL) approach for listwise LLM rerankers that eliminates the need for such training. Instead, our method relies on a small number of example rankings that demonstrate the desired trade-offs between objectives for past queries similar to the current input. We evaluate our approach on common IR test collections to investigate multiple auxiliary objectives: group fairness (TREC Fairness), polarity diversity (Touché), and topical diversity (TREC Deep Learning 2019/2020). We empirically validate that our method enables control over ranking behavior through demonstration engineering, allowing nuanced behavioral adjustments without explicit optimization. Our experiment demonstrates significant improvements in auxiliary objectives, with up to 23\% in topical diversity and 20.6\% fairness gains across different tasks, while maintaining relevance across benchmarks.

\end{abstract}

\section{Introduction}

Modern transformer-based language models are effective for ad-hoc ranking tasks~\cite{karpukhin:2020,pradeep:2023}. By learning notions of relevance from sufficient training data, these approaches often outperform unsupervised rankers \cite{karpukhin:2020,formal:2021}. However, beyond their primary objective of providing relevant content to a user, an IR system may have other auxiliary objectives, such as maximizing fairness for protected groups or topical diversity of documents \cite{Jaime_1998_novelty}. Different from relevance, which is a property of a document itself, these additional objectives, such as diversity \cite{Clarke2008} or fair representation \cite{Nick_2008}, are instead properties of a \textit{ranking}, requiring effective modeling of \textit{inter-document} interactions. 

Prior work on modeling ranking properties, such as diversity~\cite{Jaime_1998_novelty} involved interventions often in the form of ad hoc heuristic-based transformations of rankings~\cite{Rakesh_2009_diverse}. More recently, 
supervised approaches have been applied for learning neural interventions~\cite{Sean_2021_diverse} and incorporating inter-document interactions~\cite{sun2023-chatgpt}. A limitation of supervised approaches is that they are trained on a large quantity of labeled examples for each different learning objective~\cite{Schlatt2024}.

\begin{figure*}[t]
\centering
\includegraphics[width=\textwidth]{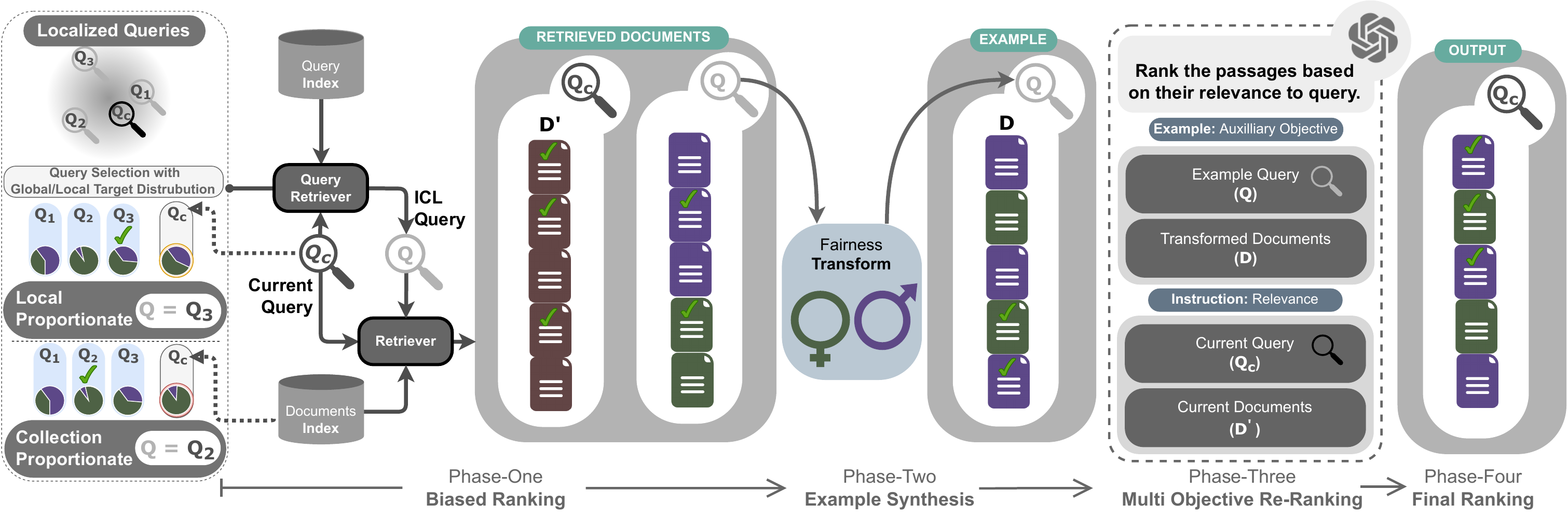}
\caption{Proposed ICL reranking method uses an example that constitutes a localized query along with its top-retrieved results arranged to satisfy a desired ranking property, such as relevance, fairness, diversity, etc. Phases Two, Three, and the selection of localized queries with local or collection proportionate, constitute our contributions.}
\label{fig:model}
\end{figure*}

To alleviate this limitation, we explore the use of instruction-tuned generative language models (LLMs) for this task, the advantage being these models can potentially act as ranking controllers adjusting their behavior through prompt examples alone~\cite{brown:2020}. While LLMs have been shown to be effective for zero-shot ranking tasks facilitated by interpreting natural language instructions~\cite{sun2023-chatgpt,pradeep:2023}, modeling list-wise objectives is more challenging because of: firstly, the difficulty of expressing them in natural language, and secondly, optimizing a prompt instruction for each different auxiliary ranking objective. Such ``prompt engineering'' is brittle~\cite{ishibashi:2023,habba:2025}, requiring iterative refinement as objectives evolve. For instance, consider a legal researcher retrieving prior legal case files to prepare counterarguments; a prompt such as ``show diverse perspectives'' may yield results that are apparently superficially varied results (e.g., lexical differences) rather than substantively balanced viewpoints (e.g., pro/con arguments on AI surveillance).

In contrast, we propose to use in-context learning (ICL)~\cite{lu:2024}, where static instructions are paired with in-context examples of a desired ranking behavior specific to a task. For instance, a single demonstration interleaving positive (pro) and negative (con) arguments could implicitly teach an LLM to diversify viewpoints, even with a task-agnostic instruction like ``order by relevance''. Our approach circumvents the need for task-specific prompts while accommodating composite or dynamic objectives by combining the primary objective (relevance) with a single auxiliary objective (diversity or fairness). As shown in Figure~\ref{fig:model}, our method reranks candidate documents using LLMs conditioned on: (1) \textit{localized (on-topic) query examples} from a large query log without requiring relevance assessments and (2) encoding ranking properties via \textit{list-wise demonstrations}, e.g., diversity and fairness. Thus, our contributions are two-fold: 

\noindent \textbf{An ICL-based approach that is shown to control desirable ranking properties without any supervised list-wise training.} 
Supervised methods rely on massive amounts of task-specific relevance training data for each new objective~\cite{Schlatt2024,Meike2020_LTR}. In contrast, our method requires no relevance judgments for the demonstration queries and requires no training. We also demonstrate that our method is flexible in handling metadata group information from the collection. For the fairness task, our collections provide explicit group metadata, which our method directly leverages. However, for the diversity task, we also demonstrate that such metadata can be derived from document clusters, even when the collection lacks explicit grouping metadata.

We establish that demonstrations can consistently change behavior across multiple objectives. Crucially, ablations confirm that demonstration construction is a key factor in effectiveness: Inverting examples significantly degrades performance, affirming its role in adapting LLM behavior. These results provide evidence for demonstration-based model adaptation as a generalizable paradigm for dynamic ranking control.

\noindent \textbf{Empirical validation of our approach, significantly improving several auxiliary objectives while maintaining ranking effectiveness.} Experiments on TREC DL (diversity)~\cite{nguyen:2016,craswell2020overview} and TREC Fairness 2022/Touché (fairness)~\cite{Ekstrand2023,bondarenko:2020} validate our approach. Unlike prior work that sacrifices ranking effectiveness for auxiliary objectives~\cite{Meike2020_LTR}, our method maintains effectiveness while significantly improving diversity and fairness. Empirical validation demonstrates that our non-parametric approach consistently outperforms heavily engineered, task-specific supervised methods (such as PAO, Set-Encoder, and DELTR) and heuristic post-hoc interventions (such as MMR and FA*IR) across four different test collections.

We provide our source code to facilitate future research and the reproducibility of our work\footnote{\url{https://github.com/nilanjansb/ICL_ranker}}.

\section{Related Work}
In this section, we review prior work on fairness and diversity in information retrieval, as well as recent advances in generative reranking and in-context learning that motivate our approach.

\para{Fairness}

Prior work has emphasized the importance of
balancing relevance with equitable exposure of document groups defined by attributes such as demographic origin, political stance, or gender~\cite{Zehlike_2017,Morik_2020}. Biased exposure in rankings risks perpetuating systemic inequities, such as amplifying majority viewpoints while suppressing marginalized perspectives \cite{Nick_2008,Ekstrand_2019}. Post-hoc fairness methods, including re-ranking algorithms~\cite{Morik_2020,Biega_2018_EOA} and fairness-aware loss functions~\cite{Singh_2018}, explicitly redistribute exposure across groups but often degrade relevance ~\cite{Pleiss_2017_fair,Sam_2017_fair}. Model-based approaches face challenges in defining target exposure distributions and incorporating them as an objective into a ranking loss~\cite{Maria_2022,Thomas_2024_FAE}.
Furthermore, supervised methods struggle with biased training data~\cite{Meike2020_LTR} and dependence on sensitive group labels, which are often incomplete or unavailable~\cite{Zehlike_2017,Thomas_2024_FAE}.%

\para{Diversity}
Different from fairness, diversity in IR addresses ambiguous queries, reducing redundancy among retrieved results \cite{DBLP:journals/tois/SenGJ22}. Diverse search results are useful in representing multiple user intents or query subtopics \cite{Jaime_1998_novelty,Clarke2008}. Similar to the fairness criteria, diversity also seeks to ensure balanced representation across various groups. However, different from fairness, the groups in diversity-based IR models correspond to facets or interpretations of a query \cite{DBLP:conf/sigir/GangulyGLJ13}.

\citet{Jaime_1998_novelty} proposed maximal marginal relevance (MMR), which is a greedy algorithm that balances between the two objectives of: i) maximizing the similarity of a document with a query, and ii) favoring documents that are dissimilar to those already ranked higher.
More recent methods infer latent query intents to synthesize diverse result sets across query intent clusters \cite{Rakesh_2009_diverse}, or employ generative models such as IntenT5 to produce a variety of plausible query interpretations \cite{Sean_2021_diverse}.
However, a limitation of these approaches is that they depend on post-hoc aggregation of intent-specific sub-rankings, which not only increases computational complexity but also often leads to a decrease in precision at top ranks \cite{Jun_2009_diverse}.

\para{Generative Rankers}
Generative rankers use autoregressive language models (LMs) to predict document permutations~\cite{sun2023-chatgpt}, bypassing traditional embedding-based or feature-centric paradigms. Recent advances in instruction-following LMs have enabled list-wise ranking, where models generate entire document orderings conditioned on a query \cite{sun2023-chatgpt,Xueguang_2024}. List-wise approaches can condition relevance on inter-document dependencies, potentially capturing subtler interactions, e.g., topic coverage, redundancy \cite{Schlatt2024}. \cite{sun2023-chatgpt} first demonstrated the effectiveness of list-wise ranking in a zero-shot setting, later adapted to smaller models via knowledge distillation \cite{pradeep:2023}. %

\para{In-Context Learning (ICL)}
ICL enables task adaptation by conditioning models on demonstration examples without parameter updates. Beyond classification and question answering \cite{Li_2023-QAICL,xu_2024-sentimentICL}, ICL allows for better out-of-distribution generalization in pairwise \cite{nilanjan2024} and list-wise ranking \cite{Xiaonan_2023_listICL}. Selecting informative examples that encode task semantics and context improves generalization~\cite{Feng_2022_ICL,nilanjan2024}.

\section{ICL for Multi-Objective Ranking}
In this section, we present a non-parametric framework for controlling list-wise ranking behavior in large language models (LLMs) using in-context learning (ICL). Our approach conditions a generative list-wise ranker on a small number of constructed example rankings, such that desired ranking properties are conveyed implicitly through demonstrated permutations rather than explicit instructions or learned parameters.

\begin{figure}[t]
\centering
\begin{adjustbox}{width=\columnwidth}
\begin{tcolorbox}[title={\small Sample target ranking from Touché dataset}]
\small
I will provide you with 3 passages, each indicated by number identifier [ ]. As an example, the first 3 passages are ranked based on their relevance to query: 
\begin{definitions*}{Similar Query from MS-MARCO}
\textbf{What is the Illinois mandate for tenured teacher evaluation?}
\end{definitions*}

\text{[1]} Teachers would be $\underbrace{\text{\bbox{limited in doing their job}}}_{\textbf{CON}}$ for ...

\text{[2]} Tenure $\underbrace{\text{\bbox{removes incentives for teachers}}}_{\textbf{CON}}$ to put in ...

\text{[3]} ... teachers get $\underbrace{\text{\gbox{ special defence from most accusations.}}}_{\textbf{PRO}}$

The three passages above are ranked based on their relevance to the search query.\\

Output: $\underbrace{\text {\bbox{[2]} > \gbox{[3]} > \bbox{[1]}}}_{\textbf{Example fair ordering}}$
\end{tcolorbox}
\end{adjustbox}
\caption{
ICL Example for a Touché query. In this example, the objective is to achieve a uniform distribution of the pro and con arguments retrieved from the Touché collection. Here, the reference query - $Q$ is lexically similar to the current input query - $Q^c$. Documents retrieved for $Q$ are reordered to balance the pro/con ratio as per the target CP or LP. This list is added to the prompt as the example output. The `PRO' and `CON' tags are not provided to the model, but are included in this diagram for better understanding.
}
\label{fig:icl_fair_examples}
\end{figure}

\para{Motivation}
Prior work has shown that ICL can effectively influence pairwise reranking~\cite{nilanjan2024}. However, many practical retrieval objectives, such as fairness and diversity, are inherently \emph{list-wise}: they depend on interactions among documents and on the distribution of attributes across a ranked list, rather than on independent document scores.

In list-wise LLM ranking, a query together with a set of retrieved documents is provided as input, and the model generates a permutation of that set \cite{sun2023-chatgpt,liu2024}. This formulation offers a natural opportunity for behavioral control, but expressing complex list-wise objectives directly as natural-language instructions is brittle and often underspecified~\citep{habba:2025}. In contrast, ICL allows ranking behavior to be shaped through examples that instantiate the desired trade-offs.

Crucially, this control mechanism is \emph{non-parametric}: it does not require training or fine-tuning a model for each auxiliary objective. Instead, control is achieved by modifying only the example rankings supplied at inference time, enabling flexible adaptation to different ranking criteria without additional supervision.

\para{Localised Example Selection}

The effectiveness of ICL in ad-hoc ranking has been shown to depend on selecting demonstrations that are topically aligned with the test query while remaining free of relevance labels~\citep{nilanjan2024}. Given a test query $Q_c$, the first step in our proposed workflow is to use a relatively large repository of existing queries $\mathcal{Q}$ to retrieve a set of $k$ most \textit{similar queries} - $\mathcal{N}_k(Q_c)$ (see Figure~\ref{fig:loc_query_examples} of Appendix~\ref{ss:prompt-template}). For our experiments, we use the MS MARCO training set as our query reference set $\mathcal{Q}$ \textit{without} labeled examples. We use BM25 over the MS~MARCO training queries and a dense index of BERT vectors to obtain \(\mathcal{N}_k(Q_c)\). We refer to the lexical and semantic example selections as \textbf{LEX} and \textbf{SEM}, respectively. From the retrieved queries, the query with the ranked list attribute distribution closest to the target distribution is selected (one demonstration query), shown as ``Query Retriever'' in Figure~\ref{fig:model}. We refer to this selected demonstration query as \(Q\). Details on the exploration of other higher values of \(k\) are provided in the Appendix \ref{appendix:query_selection}.

\para{Implicit multi-objective control}
Our method encodes auxiliary objectives entirely through the \emph{structure} of the example ranking. The prompt instruction remains fixed (“rank by relevance”), while the example permutation exhibits a controlled distribution over document attributes. The example is relevance-ordered within attribute groups, but its composition approximates a target distribution $\tau$.

As a result, the model is never explicitly instructed to optimize for fairness or diversity. Instead, it conditions on a relevance-consistent demonstration that implicitly specifies how relevance should be traded off against other ranking properties when generating a permutation for the test query $Q_c$.

We then construct an \textbf{example ranking} to be used as additional context to condition the generative output of an LLM. To obtain this ICL example ranking, the top-retrieved query (i.e., $Q$) is used to retrieve a set of top-$m$ ranked documents from a target collection, say $\mathcal{T}$. We denote this top-retrieved list as $\theta(Q)_m = \{D_1,\ldots, D_m\}$, where $\theta$ is a retrieval model. This example ranking is constructed offline and serves solely as contextual input to the LLM at inference time.

The next step is to induce an ordering on the set of top-$m$ documents $\theta(Q)_m$, i.e., $\theta(Q)_m \mapsto \langle D_{\pi(1)},\ldots, D_{\pi(m)}\rangle$, where $\langle \cdot \rangle$ represents a sequence, and $\pi$ denotes a permutation function over sets of $m$ elements.
In the simplest case, the permutation function $\pi$ corresponds only to the primary objective of relevance, i.e., maximizing the likelihood of positioning a relevant document ahead of a non-relevant one in the sequence. In practice, such information is available for a training set of queries (e.g., the MS MARCO train set), and prior work has shown that the inclusion of these examples improves ranking preferences \cite{nilanjan2024}. In addition to relevance, this permutation function can be designed to correspond to another auxiliary task, thus allowing provision for a more general use case, which we discuss next.

\para{Target Distribution-based Ranking}
The auxiliary objective is specified as a target distribution over categorical document attributes, rather than as an explicit optimization instruction. The example ranking manifests this distribution while preserving within-group relevance order, thereby inducing a list-wise preference structure under a single relevance prompt.
More formally, $\forall D \in \topdocs{Q}$, let $A(D) \in \mathbb{Z}_k$ denote the value of the attribute $A$ for document $D$. For instance, $A$ may refer to the gender of an entity within a document, in which case, $k=2$ and $A(D) \in \{\text{`Male'}, \text{`Female'}\}$. Conversely, for topical diversity tasks, $A(D)$ maps to $k$ distinct aspects (where $k > 2$), representing a multi-group distribution across diverse query interpretations.

For an IR task with an auxiliary objective involving an attribute $A$, a target distribution of these metadata values over the set of relevant documents is specified as a part of the input.
For a query $Q$, we denote this distribution as $\tau({\qrel{Q}}) \in \mathbb{R}^k$ (where $k$ is the number of possible values). The $i^{th}$ component of this vector is given by
\begin{equation}
\tau({\qrel{Q}_i}) = \frac{\sum_{D \in \qrel{Q}}\mathbb{I}(A(D) = i)}{|\mathcal{R}(Q)|},   \label{eq:tgt} 
\end{equation}
where $(1 \leq i \leq k)$, which represents the relative proportion of relevant documents for each metadata value ($\mathbb{I}$ denotes the indicator function).

Our approach is flexible with respect to any number of $k$. In fairness tasks, group selection is typically binary (since the chosen attributes are generally limited in such contexts), whereas in diversity tasks, multiple groups are considered.

\para{Operational source of \(\tau\)} In realistic deployment settings, relevance judgments are unavailable at test time; hence \(\tau(\mathcal{R}(Q))\) of Equation \ref{eq:tgt} must be instantiated without using test relevance labels. Throughout our experiments, we therefore obtain \(\tau\) from signals available \emph{prior} to re-ranking: (i) for fairness, we use benchmark-provided document attributes or corpus-level priors; (ii) for diversity, we assume a uniform target over automatically induced aspect clusters (see Appendix \ref{appendix:clustering}). This avoids train–test leakage while keeping the method non-parametric. Since these attributes are part of the collection, we refer to them as Collection Proportionate (\textbf{CP}). It ensures that the example ranking matches the distribution of the collection, preventing a biased reranking. In addition to global priors, we also define a target distribution that accommodates query-specific constraints. Specifically, we define Local Proportionate (\textbf{LP}), which dynamically derives the target distribution from the attribute distribution within the top-retrieved candidate list of the test query.

\para{Grouping by metadata values}
We aim to rerank \(\theta(Q)_m\) so that prefixes approximate \(\tau\). Treating retrieval scores as \emph{monotonic surrogates} for relevance\footnote{Our approach inherits biases of the first-stage ranker, since example construction assumes monotonic relevance ordering within groups.}, we partition \(\theta(Q)_m\) into \(k\) sequences \(\{\topdocs{Q}^{(i)}\}_{i=1}^k\) by attribute value \(A(D)=i\), preserving the within-group retrieval order. We do not expose labels to the ranker at runtime, either in demonstrations or test instances. 

\para{Auxiliary Objective based Rank Induction}
To instantiate the target distribution while minimally perturbing relevance, we construct the example ranking using a greedy prefix-based selection procedure. Let \(p\) denote the current prefix length and let \(\mathcal{C}_{p+1}\) contain the next unselected candidate (head) from each group \(\topdocs{Q}^{(i)}\). We greedily select the \((p{+}1)\)-th document to minimize the divergence between the prefix histogram and the target:
\begin{equation}
\label{eq:selection}
\begin{split}
D_{p+1} = \arg\min_{D\in\mathcal{C}_{p+1}} \mathrm{KL} \bigl( &\tau(\mathcal{R}(Q)), \\
&\tau(\langle D_1,\ldots,D_p\rangle \cup D) \bigr).
\end{split}
\end{equation}
After selection from group \(j\), we advance that group’s pointer and repeat until \(m\) items are chosen. The procedure defined in Equation \ref{eq:selection} mirrors classical exposure-control heuristics, but is used here solely to synthesize a demonstration rather than to directly rerank test documents. Demonstrations are then appended before the test instance to influence ranking behaviour.

\section{Evaluation}
We now provide empirical evidence for our approach, structured around three research questions.

\noindent\textbf{RQ-1}: Can example rankings improve effectiveness in a zero-shot list-wise ranking task?

\noindent \textbf{RQ-2}: How do ICL examples compare to directly instructing an LLM in terms of auxiliary objective effectiveness?

\noindent \textbf{RQ-3}: What are the effects of ICL example ranking strategies that are adversarial to a target distribution-based auxiliary objective?

\subsection{Experimental Setup}

\para{Objectives and Datasets}
Our methodology is fundamentally task-agnostic and designed to generalize to any list-wise reranking objective. However, our current investigation focuses on two auxiliary objectives: diversity and group fairness. To improve diversity, we aim to retrieve relevant but topically diverse content to satisfy multiple potential information needs under ambiguity. To operationalize this, we adopt the experimental framework of \citet{Schlatt2024} using the MSMARCO passage corpus~\cite{nguyen:2016} and TREC Deep Learning (\textbf{DL-2019} and \textbf{DL-2020})~\cite{craswell2020overview,craswell2021overview} test collections.  In contrast to fairness, where group labels in relevance judgments inform the target distribution, diversity assumes a uniform distribution over latent topics, derived via clustering over retrieved documents. Additional details on the clustering procedure are given in Appendix \ref{appendix:clustering}, and full dataset descriptions are provided in Appendix \ref{appendix:dataset}. We also evaluate group fairness in the single-ranking setting using TREC \textbf{Fair-2022}~\cite{Ekstrand2023} and \textbf{Touché}~\cite{bondarenko:2020}, both of which include explicit group attributes. %

Our method allows a query log from any collection to serve as an anchor set. To demonstrate this flexibility, we use a query reference set from the MSMARCO training set (approximately $8 \times 10^5$ entries) to construct demonstrations for the in-domain test sets DL-2019 and DL-2020. We then use the same query reference set, without any modification, to construct demonstrations for entirely distinct, out-of-domain test collections such as Fair-2022 and Touché.

\para{Measures}
To evaluate ranking quality, we report Normalized Discounted Cumulative Gain (\textbf{nDCG}) for relevance as is standard on chosen benchmarks~\cite{craswell2020overview,craswell2021overview,bondarenko:2020,Ekstrand2023}. For measuring fairness, we apply Attention-Weighted Rank Fairness (\textbf{AWRF}) and nDCG-AWRF combination (\textbf{M1}), as is standard in group fairness~\cite{Ekstrand2023}. We report \textbf{$\alpha$nDCG} following~\citet{Clarke2008} with $\alpha=1$ to measure diversity (see Section \ref{sec:ablation_alpha}). All metrics are reported at a cutoff of 10.

\para{Multiple Comparison Correction}
To test the statistical significance of our results, we apply the Holm-Bonferroni correction to account for multiple hypothesis testing. Specifically, we define separate comparison families based on the reference baseline. We compare against the first-stage retrieval baseline (BM25/ColBERT) and against the 0-shot baseline. The correction is applied jointly within each family across all evaluated tasks, datasets, retrievers, and metrics.

\para{Configurations}
Our ICL methods are labeled to denote the explicit components of our demonstration engineering setup, comprising of \textbf{Diverse} or \textbf{Fair} objectives, \textbf{LEX} or \textbf{SEM} similarity, and \textbf{CP} or \textbf{LP}-based target distribution. For example, a pipeline defined as Diverse-LEX-CP indicates an auxiliary objective of topical diversity, using a lexical similarity-based retrieval mechanism to select a localized query from the query log, and mapping the target distribution ($\tau$) using global collection metadata priors.

\para{Models}
We re-rank the top 100 documents using a two-stage pipeline. We apply BM25~\cite{robertson:1994} and ColBERT~\cite{colbert-sigir2020} as initial retrieval models. The second-stage ranker is our proposed ICL-based method. We apply a sliding window-based re-ranking over the top 100 documents. The window size is fixed at 20 documents with a stride size of 10~\cite{sun2023-chatgpt,pradeep:2023}. 

We perform all experiments using \texttt{GPT-4o-mini} in the main experiments. We also test the effectiveness of our approach on different models, comparing the closed-source \texttt{GPT-4o-mini} with the open-weighted \texttt{Llama-3.1 (8B \& 70B)} described in detail in Appendix \ref{appendix:llm_comparison}. Details on model configurations, hyper-parameters, and implementation are provided in Appendix \ref{appendix:setup}.

\para{Ablations}

Our ablation aims to establish that our augmentation of ICL target distributions indeed leads to improved effectiveness; thus, we ablate how we construct list-wise demonstrations. \textit{Adversarial Examples} (\textbf{Adv}): By swapping the proportions, we seek to maximize the KL divergence from the target distribution (instead of minimizing this measure as per Equation \ref{eq:selection}), e.g., flipping a 3:2 gender ratio to 2:3. \textit{Relevant Examples} (\textbf{Rel}): A transformation from a random order to an order by relevance determined by a first-stage system. \textit{Static Examples} (\textbf{Static}): Replace similar query examples with a fixed example ranking used for all test queries, which implies no shared topicality. The purpose is to explore whether topicality is important for effective ICL example rankings.

\subsection{Baselines} \label{sec:baselines}
\para{Prompt-based Auxiliary Objective (PAO)}
This is a 0-shot list-wise ranking baseline that uses explicit instructions to elicit fairness or diversity from the model. The prompt for fairness is: 
``\textit{Rank the passages based on their fairness, ensuring that ranked results do not discriminate against certain individuals, groups, or entities}'', whereas for diversity, it is: 
``\textit{Rank the passages based on their topical diversity, ensuring that ranked results contribute to different topics uniformly}''. The base model applied for PAO and our approach is identical (GPT is shown in our primary findings).

\para{Baselines for Diversity}
For the diversity-ranking task, we compare our approach to the following baselines.

\noindent \textbf{Max Margin Relevance (MMR)}~\cite{Jaime_1998_novelty}: Combines relevance and diversity via a linear combination. Following \citet{Yadong_2014-diverse}, the mixture weight was tuned with 5-fold cross-validation.

\noindent \textbf{Set-Encoder}~\cite{Schlatt2024}: A cross-encoder trained with diversity-aware loss on MSMARCO. We use \texttt{set-encoder-large}\footnote{\url{https://github.com/webis-de/set-encoder}}.

\para{Baselines for Fairness} For the fair-ranking task, we compare our approach the following.
\noindent \textbf{FA*IR}\footnote{\url{https://github.com/fair-search/fairsearch-fair-python}}~\cite{Zehlike_2017}: A post-processing method to enforce fair exposure. The two hyperparameters of FA*IR, namely a) $\alpha$: the proportion of protected candidates, and b) $p$: the significance level, were tuned via 10-fold cross-validation over $\alpha \in [0.01,0.13]$ and $p \in [0.4,0.85]$ with step sizes of $0.01$. The optimal values were found to be $(\alpha, p) = (0.1, 0.2)$.
        
\noindent \textbf{DELTR}\footnote{\url{https://github.com/fair-search/fairsearch-deltr-python}}~\cite{Meike2020_LTR}: A learning-to-rank model trained on fair rankings created using FA*IR. We use our example sets to simulate fair supervision. We set a high value on the trade-off parameter $\gamma$=1, to ensure that it promotes fairness without impacting the overall ranking utility \cite{Meike2020_LTR}.

\subsection{Findings}

\begin{table}[t]
\centering
\caption{
Evaluating nDCG and $\alpha$nDCG performance over TREC DL-2019 and 2020 for the Diversity objective. The maximum score in each category is shown in bold. Symbols $\star$ and $\dagger$ indicate statistical significance after Holm–Bonferroni correction (adjusted $p < 0.05$) for comparisons against the first-stage and 0-shot baselines, respectively.
\label{tab:nov100_results}
}
\begin{adjustbox}{width=\columnwidth}
\begin{tabular}{@{}l@{~~}l@{~~}l@{~~}l@{~~}l@{~~}l@{~~}l@{}}
\toprule
& & & \multicolumn{2}{c}{DL-2019} & \multicolumn{2}{c}{DL-2020} \\
\cmidrule(r){4-5}
\cmidrule(r){6-7}
Type & & Pipeline & nDCG & $\alpha$nDCG & nDCG & $\alpha$nDCG \\
\midrule
\multirow{5}{*}{\rotatebox[origin=c]{90}{Baselines}}& \textit{1} & BM25 & .4795 & .4569 & .4936 & .4895\\
& \textit{2} & + PAO & .6817 & .6844 & .6349 & .6670\\
& \textit{3} & + MMR & .4786 & .4559 &  .4922 & .4899  \\
&\textit{4} & + Set-Encoder & .5977 & .5695 & .5971 &  .6357\\
& \textit{5} & + 0-shot & .6971 & .6761 & .6826 &  .7039\\
\cmidrule(r){3-7}
\multirow{4}{*}{\rotatebox[origin=c]{90}{ICL}} & \textit{6} & +  Diverse-LEX-CP & $\textbf{.7124}^{\star}$ & $.7135^{\star}$ & $\textbf{.6844}^{\star}$ &  $.7228^{\star}$\\
& \textit{7} & + Diverse-SEM-CP & $.7109^{\star}$ & $\textbf{.7309}^{\star \dagger}$ & $.6831^{\star}$ &  $\textbf{.7355}^{\star}$\\
& \textit{8} & + Diverse-LEX-LP & $.6977^{\star}$ & $.7098^{\star}$ & $.6814^{\star}$ &  $.7141^{\star}$\\
& \textit{9} & + Diverse-SEM-LP & $.7089^{\star}$ & $.7254^{\star \dagger}$ & $.6824^{\star}$ &  $.7223^{\star}$\\

\midrule
\multirow{5}{*}{\rotatebox[origin=c]{90}{Baselines}}& \textit{10} & ColBERT & .7205 & .6583 & .6864 & .6385\\
& \textit{11}& + PAO & .6949 & .6494 & .6996 & .6910\\
& \textit{12}& + MMR & .7173 & .6567 & .6873 & .6405\\
& \textit{13}& + Set-Encoder & .7320 & .6172 & .7245 & .6338\\
& \textit{14}& + 0-shot & \textbf{.7699} & .6850 & .7498 & .6843\\
\cmidrule(r){3-7}
\multirow{4}{*}{\rotatebox[origin=c]{90}{ICL}} & \textit{15} & + Diverse-LEX-CP & .7601 & .6891 & $\textbf{.7700}^{\star \dagger}$ &  $.7132^{\star}$ \\
& \textit{16} & + Diverse-SEM-CP & $.7476$ & $\textbf{.7012}^{\star}$ & $.7543^{\star}$ &  $\textbf{.7322}^{\star \dagger}$\\
& \textit{17} & + Diverse-LEX-LP & $.7582$ & $.6827$ & $.7638^{\star}$ &  $.7090^{\star}$\\
& \textit{18} & + Diverse-SEM-LP & $.7419$ & $.6833$ & $.7598^{\star}$ &  $.7104^{\star}$\\
\bottomrule
\end{tabular}
\end{adjustbox}
\end{table}

\begin{table}[t]
\centering
\caption{
Evaluating AWRF, nDCG, and M1 on Touché-2020 (PRO/CON) and TREC Fair-2022 (M/F) for the fairness objective; other details are similar to Table \ref{tab:nov100_results}.}
\label{tab:fair100_results}
\begin{adjustbox}{width=\columnwidth}
\begin{tabular}{@{}l@{~}l@{~~}l@{~~}l@{~~}l@{~~}l@{~~}l@{~~}l@{~~}l@{}}
\toprule
 & & & \multicolumn{3}{c}{Touché-2020} & \multicolumn{3}{c}{Fair-2022} \\ 
 \cmidrule(r){4-6}
\cmidrule(r){7-9}
Type & & Pipeline & nDCG & AWRF & M1 & nDCG & AWRF & M1 \\ \midrule
\multirow{5}{*}{\rotatebox[origin=c]{90}{Baselines}} & \textit{1} &BM25 & .2530 & .4811 & .1851 & .4974 & .4901 & .2975 \\
 & \textit{2} & + PAO & .2258 & .5218 & .1589 & .5667 & .5312 & .3332 \\
 & \textit{3} & + FA*IR & .2452 & .4620 & .1660 & .3735 & \textbf{.7215} & \textbf{.3989} \\
 & \textit{4} & + DELTR & .2486 & .3212 & .1190 & .3786 & .5593 & .3220 \\
 & \textit{5} & + 0-shot & .2590 & .5377 & .1936 & .5688 & .5494 & .3428 \\
 \cmidrule(l){3-9} 
\multirow{4}{*}{\rotatebox[origin=c]{90}{ICL}} & \textit{6} & + Fair-LEX-CP  & \textbf{.2608} & $\textbf{.5800}^{\star \dagger}$ & $\textbf{.2023}$ & $\textbf{.5697}^{\star}$ & $.5697^{\star}$ & $.3526^{\star}$ \\  
& \textit{7} & + Fair-SEM-CP  & .2531 & $.5594^{\star}$ & .1960 & $.5637^{\star}$ & $.5549^{\star}$ & $.3485^{\star}$ \\
& \textit{8} & + Fair-LEX-LP  & .2516 & $.5437^{\star}$ & .1949 & $.5600^{\star}$ & $.5484^{\star}$ & $.3416^{\star}$ \\
& \textit{9} & + Fair-SEM-LP  & .2308 & $.5420^{\star}$ & .1912 & $.5596^{\star}$ & $.5488^{\star}$ & $.3418^{\star}$ \\

\midrule
\multirow{5}{*}{\rotatebox[origin=c]{90}{Baselines}} & \textit{10} & ColBERT & \textbf{.2590} & .2994 & .1462 & .4854 & .2068 & .1204 \\
 & \textit{11} & + PAO & .2234 & .2388 & .0956 & .6565 & .1837 & .1411 \\
 & \textit{12} & + FA*IR & .2500 & \textbf{.3598} & \textbf{.1698} & .2111 & \textbf{.5896} & \textbf{.3320} \\
 & \textit{13} & + DELTR & .2518 & .2216 & .1078 & .2128 & .3370 & .2215 \\
 & \textit{14} & + 0-shot & .2496 & .2197 & .1027 & .6487 & .2056 & .1466 \\
 \cmidrule(l){3-9} 
 \multirow{4}{*}{\rotatebox[origin=c]{90}{ICL}} & \textit{15} & + Fair-LEX-CP & .2508 & $.2602^{\dagger}$ & .1216 & $\textbf{.6606}^{\star}$ & .2272 & $.1628^{\star}$ \\
 & \textit{16} & + Fair-SEM-CP  & .2503 & .2316 & .1137 & $.6527^{\star}$ & .2148 & $.1522^{\star}$ \\
& \textit{17} & + Fair-LEX-LP  & .2576 & .2318 & .1142 & $.6502^{\star}$ & .2011 & .1467 \\
& \textit{18} & + Fair-SEM-LP  & .2545 & .2303 & .1139 & $.6513^{\star}$ & .2117 & $.1516^{\star}$ \\
\bottomrule
\end{tabular}
\end{adjustbox}
\end{table}

In this section, we present and discuss our findings. 

\para{Examples allow for the modeling of multiple objectives} Our experiments demonstrate that in-context learning (ICL) with task-guided examples enables effective optimization of auxiliary objectives while maintaining relevance (\textbf{RQ-1}). For diversity modeling (Table \ref{tab:nov100_results}), our approaches significantly outperform the first stage and 0-shot baselines in topical diversity ($\alpha$nDCG), with improvements of up to 23\% (e.g., compare rows 1,5 with 6-9 and 10,14 with 15-18). We observe that semantic query retrieval (SEM) methods are more effective in-domain, aligning with earlier findings~\cite{nilanjan2024}. Notably, these surpass both post-hoc diversification (MMR) and a supervised list-wise method (Set-Encoder) by 8-15\% in $\alpha$nDCG (compare rows 3-4 with 6-9 and 12-13 with 15-18), despite relying only on heuristic examples rather than explicit optimization. These results indicate that example-based task conditioning provides a sufficient learning signal for the model to acquire objective-specific ranking behaviors.

Table \ref{tab:fair100_results} shows results for the auxiliary objective of group fairness. Our approach exceeds all baselines on Touché and is competitive with post-processing and supervised methods on Fair-2022. Our method achieves approximately a 52\% improvement in nDCG performance compared to DELTR and FA*IR applied post-hoc to an LLM relevance ranking, as shown by comparing Rows 3-4,12-13 with 6-9,15-18 in Table~\ref{tab:fair100_results}. However, this gain in effectiveness is accompanied by a reduction in fairness on the Fair-2022 dataset. For this task, however, we observe that lexical query retrieval is more effective out-of-domain, consistent with existing findings~\cite{nilanjan2024}. Table~\ref{tab:fair100_results} also reveals a sensitivity to the choice of first-stage retriever: BM25 yields better fairness outcomes than ColBERT across evaluation settings. Notably, under the ColBERT retriever, our approach shows diminished effectiveness on Touché. This trend aligns with prior findings~\cite{tang:2023,parry:2024}, highlighting that weaker first-stage rankings tend to impair the effectiveness of list-wise re-ranking. Interestingly, under a diversity-oriented objective, we observe the opposite pattern: stronger first-stage rankings result in reduced diversity effectiveness. We analyze these interactions, particularly the role of positional bias, in detail in Appendix \ref{appendix:ex_ordering}.

Overall, our methods using collection proportionate (CP) are consistently better than those using local proportionate (LP). This, however, is not a drawback; instead, it indicates reliance on the distribution of collection attributes in evaluation measures (such as AWRF). Additionally, this highlights the flexibility of our approach, which adapts to both global and local distributions to select localized queries for ICL, advancing over existing ICL approaches for context selection~\cite{nilanjan2024,Xie_2024-ICL,Li_2023-QAICL}.

Nevertheless, our approach with ICL outperforms the 0-shot setting. With BM25 as the first-stage ranker, our approach is more effective than all but one baseline, FA*IR with grid-search optimised parameters, on Fair-2022. Under ColBERT, FA*IR is most effective across both datasets. However, our approach outperforms all other baselines but is at near-parity with 0-shot, suggesting a minimal change in the model's process. The FA*IR approach requires prior group information for ranking and poses a significant trade-off in relevance, thereby limiting its practical applicability.%

\begin{table}[t]
\centering
\caption{
Ablations showing effects on diversity and fairness tasks. The ``+Target'' row corresponds to the target-distribution specific ICL re-ranking results  using LEX-CP.
Suffixes \textit{a} to \textit{e} represent the statistical significance of ``+Target'' when compared to the first-stage, 0-shot, Adv, Rel, and Static methods, respectively, computed via Holm–Bonferroni correction (adjusted $p < 0.05$).
\label{tab:nov100_ablation_results}
}
\begin{adjustbox}{width=\columnwidth}
\begin{tabular}{@{}l@{~~}l@{~~}l@{~~}l@{~~}l@{~~}l@{~~}l@{~~}l@{~~}l@{}}
\toprule
& \multicolumn{2}{c}{DL-2019} & \multicolumn{2}{c}{DL-2020} & \multicolumn{2}{c}{Touché-2020} & \multicolumn{2}{c}{Fair-2022}\\
\cmidrule(r){2-3}
\cmidrule(r){4-5}
\cmidrule(r){6-7}
\cmidrule(r){8-9}
Method & nDCG & $\alpha$nDCG & nDCG & $\alpha$nDCG & nDCG & AWRF & nDCG & AWRF \\
\midrule
BM25 & .479 & .457 & .494 & .489 & .253 & .481 & .497 & .490 \\
+ 0-shot & .697 & .676 & .683 &  .704 & .259 & .537 & .569 & .549 \\
\cmidrule(r){2-9}
\rowcolor{lightgray}
+Target & $\textbf{.712}^{\textit{a}}$ & $\textbf{.713}^{\textit{abe}}$ & $.684^{\textit{a}}$ &  $\textbf{.729}^{\textit{ac}}$ & \textbf{.261} & $\textbf{.580}^{\textit{abcde}}$ & .570 & $\textbf{.570}^{\textit{a}}$  \\
+Adv & .696 & .688 & .686 & .701 & .260 & .520 & .572 & .550 \\
+Rel & .700 & .685 & \textbf{.692} & .712 & .255 & .524 & \textbf{.581} & .548 \\
+Static & .704 & .680 & .682 & .709 & .251 & .480 & .570 & .550\\
\midrule
ColBERT & .720 & .658 & .686 & .638 & \textbf{.259} & \textbf{.299} & .485 & .207\\
+0-shot & \textbf{.770} & .685 & .750 & .684 & .250 & .220 & .649 & .206 \\
\cmidrule(r){2-9}
\rowcolor{lightgray}
+Target & .760 & $\textbf{.689}^{\textit{ac}}$ & \textbf{.770} &  $\textbf{.713}^{\textit{abe}}$ & .251 & $.260^{\textit{bcde}}$ & \textbf{.661} & $\textbf{.227}^{\textit{ce}}$ \\
+Adv & \textbf{.770} & .686 & .761 & .706 & .253 & .197 & .656 & .185  \\
+Rel & .764 & .682 & .762 & .710 & .245 & .217  & .646 & .185 \\
+Static & .760 & .687 & .755 & .698 & .241 & .181 & .648 & .185 \\
\bottomrule
\end{tabular}
\end{adjustbox}
\end{table}

\para{In-context learning improves ranking performance over direct instructions}
We now examine the impact of prompt-based optimization on fairness and diversity objectives(\textbf{RQ-2}). As shown in Tables~\ref{tab:nov100_results} and~\ref{tab:fair100_results}, our ICL approach is more effective than the Prompt-based Auxiliary Objective (PAO) baseline. Relative to the 0-shot baseline, the PAO method exhibits a notable decline in nDCG, with the sole exception occurring under the ColBERT retriever on the Fair-2022 dataset. This reduction in ranking utility is likely attributable to the modified prompt, which explicitly instructs the model to enhance diversity--an objective combination that may be out-of-distribution for the model. This observation further underscores the advantages of demonstration-based adaptation. Marginal improvements are found in $\alpha$nDCG with PAO on DL-2019 under BM25 and on DL-2020 under ColBERT (Table~\ref{tab:nov100_results}, Rows 2 and 11). In terms of AWRF, a positive gain is observed only for Touché when ColBERT is used as the first-stage retriever (Table~\ref{tab:fair100_results}, Row 11). We attribute the inconsistency in PAO performance to the lack of explicit contextual grounding regarding the fairness or diversity criteria being targeted, as opposed to the model's default interpretation of these objectives.

\para{Example inversion largely degrades effectiveness}
From Table \ref{tab:nov100_ablation_results}, we see that examples optimizing directly for relevance yield significant improvements in nDCG relative to both BM25 and ColBERT. This is a natural consequence of the original motivation for ICL demonstrations~\cite{lu:2024}. From Rows 4 and 10, see a consistent degradation in terms of $\andcg$ when applying a uniform example (the adversarial setting for the diversity task). This suggests that the target distribution-induced ordering plays an important role in ranking  (\textbf{RQ-3}).

Additionally, see that ICL examples exhibit minimal trade-off in terms of relevance, suggesting that fine-grained control can be exerted without compromising the core task except for static examples as observed from Rows 6 and 12 of Table \ref{tab:nov100_ablation_results}. Static examples lead to a substantial decline in effectiveness, occasionally reducing it below that of the base ranker. This validates that the locality of queries in ICL examples remains useful in a list-wise setting, as was found by \citet{nilanjan2024} in a pair-wise setting.

\begin{table}[t]
\centering
\small
\caption{Sensitivity to the $\alpha$ parameter on DL-2019 and DL-2020. We compare first-stage retrieval (BM25 and ColBERT), 0-shot, and ICL (Diverse-LEX-CP) across $\alpha \in \{0.5, 0.75, 1\}$. We use bold font to denote $\alpha=1$, our primary setting.}
\label{tab:ablation_alpha}
\begin{adjustbox}{width=0.9\columnwidth}
\begin{tabular}{l@{\hspace{3pt}}r@{\hspace{2pt}}cccccc}
\toprule
\multirow{2}{*}{} & & \multicolumn{3}{c}{DL-2019} & \multicolumn{3}{c}{DL-2020} \\ \cmidrule(l){3-5} \cmidrule(l){6-8}
Model&$\alpha$& \textbf{1} & $0.75$ & $0.5$ &\textbf{1} & $0.75$ & $0.5$ \\ \midrule
BM25 & & .457 & .469 & .480 & .490 & .491 & .492 \\
0-shot&  & .676 & .689 & .700 & .704 & .721 & .746 \\
ICL& & .714 & .726 & .739 & .723 & .743 & .757 \\ \midrule
ColBERT& & .658 & .681 & .705 & .639 & .665 & .671 \\
0-shot& & .685 & .692 & .697 & .684 & .695 & .706 \\
ICL& & .689 & .694 & .698 & .713 & .718 & .722 \\ \bottomrule
\end{tabular}
\end{adjustbox}
\end{table}

\para{Sensitivity Analysis of $\alpha$nDCG}
\label{sec:ablation_alpha}
To assess the sensitivity of our results to the choice of $\alpha$ in $\alpha$nDCG, we evaluate three settings, $\alpha \in \{0.5, 0.75, 1\}$. While $\alpha=0.5$ is a commonly used convention, our primary setting of $\alpha=1$ sets high emphasis on diversity in the relevance--diversity trade-off. This choice is motivated by our objective of evaluating the model's capacity to diversify, while we evaluate relevance separately using standard nDCG.

To ensure that our findings are not an artifact of the $\alpha=1$ setting, we compare first-stage retrieval, 0-shot, and ICL across different $\alpha$ values. As shown in Table~\ref{tab:ablation_alpha}, ICL consistently outperforms the corresponding 0-shot setting for both datasets across different values of $\alpha$. These consistent gains demonstrate that the effectiveness of our proposed ICL method is robust to the choice of $\alpha$. We use $\alpha=1$ as our primary setting because it places greater emphasis on the diversification capability that is central to our task.

\para{Concluding Remarks}
We propose an approach to multi-objective ranking that leverages demonstrations to balance relevance and auxiliary objectives. Our experiments confirm that examples modeled for fairness and diversity improve target objectives significantly without compromising relevance. We found that effectiveness gains can be controlled with adversarial examples, leading to biased downstream rankings. Additionally, our approach improves over directly instructing a model for each objective. Our approach demonstrates greater effectiveness compared to task-specific post-hoc and supervised methods, while effectively mitigating the potential trade-offs and the need for task-specific modifications. Furthermore, our findings present encouraging evidence for demonstration-based model adaptation as a mechanism for controlling ranking behavior beyond the objectives investigated in this work. Our ICL-based approach for ranking topical diversity achieves gains of up to 23\% in the $\alpha$nDCG metric and a 20.6\% improvement in the AWRF metric for fairness, while often preserving or improving nDCG.

\section{Limitations}
We do not explore all possible avenues for demonstration-based multi-objective search in this work. Indeed, several parameter choices are motivated by prior work; however, due to our novel setting, it could be that under this new setting, effectiveness could be further improved.

We demonstrate that a single example is generally sufficient for the collections in our experiments. We expect modern LLMs with large context windows to accommodate applications involving longer documents, such as legal or medical retrieval. However, we acknowledge that complex real-world applications may benefit from additional examples that may pose greater computational and representational challenges.

We represent gender as binary in accordance with the collection metadata; however, this simplification does not fully capture the complexities of gender in the real world.

Our approach requires an existing query log, which, in low-information environments or low-resource languages, may pose difficulties for adoption. In future work, we look to rectify the need for a monolingual corpus.

\bibliography{bibliography}

\appendix

\section{Additional Experimental Details}
\label{appendix:setup}

\subsection{Dataset Description}
\label{appendix:dataset}

The fairness task corresponds to that of the `single ranking' task of TREC Fairness track \cite{Ekstrand2023} on the `eval' query set. The objective in the Touché task is to mitigate the bias towards a specific stance \cite{DBLP:journals/corr/KulshresthaEMZG17}, whereas the objective in the ad-hoc search task on TREC DL topics is to maximize the topical diversity, where each topic maps to a cluster of documents. We illustrate the details of our chosen collections in Table~\ref{tab:dataset_desc}.

\begin{table}[t]
\centering
\caption{
Statistics of the datasets used in our experiments.
}
\label{tab:dataset_desc}
\begin{adjustbox}{width=\columnwidth}
\setlength{\tabcolsep}{3pt}
\small
\begin{tabular}{llccccc}
\toprule
Task & Collection & $|\mathcal{C}|$ & Queries & $|\mathcal{Q}|$ & Name & Values \\
\midrule
\multirow{2}{*}{Fairness} & TREC Fair & 6.5M & Fair-2022 & 50 & Gender & M, F\\
& TouchéV2 & 383K & Touché'20 & 49 & Stance & PRO, CON \\
\cmidrule{1-7}
\multirow{2}{*}{Diversity} & \multirow{2}{*}{MS MARCO} & \multirow{2}{*}{8.8M} & DL'19 & 43 & \multirow{2}{*}{Topic} & \multirow{2}{*}{$\mathbb{Z}$}  \\
& & &DL'20 & 54 & & \\
\bottomrule
\end{tabular}
\end{adjustbox}
\end{table}

\subsection{Clustering for Diversity}
\label{appendix:clustering}
To induce topic clusters for diversity evaluation, we apply hierarchical agglomerative clustering with complete linkage over Jaccard similarity between token sets. This is applied to the top 100 documents retrieved per query. Due to the nature of agglomerative clustering, the number of clusters varies with query specificity, resulting in a query-dependent target distribution.

\subsection{Query Similarity Retrieval}
Following~\cite{nilanjan2024}, we retrieve similar queries from the MSMARCO training split using BM25 over query text. We retrieve the top-3 most similar queries and aggregate their retrieved documents to build example sets for in-context learning. No relevance judgments are used in this process.

\subsection{Number of Examples}
\label{appendix:query_selection}

\begin{figure}[h]
    \centering
    \begin{subfigure}[b]{0.8\columnwidth}
        \centering
        \includegraphics[width=\columnwidth]{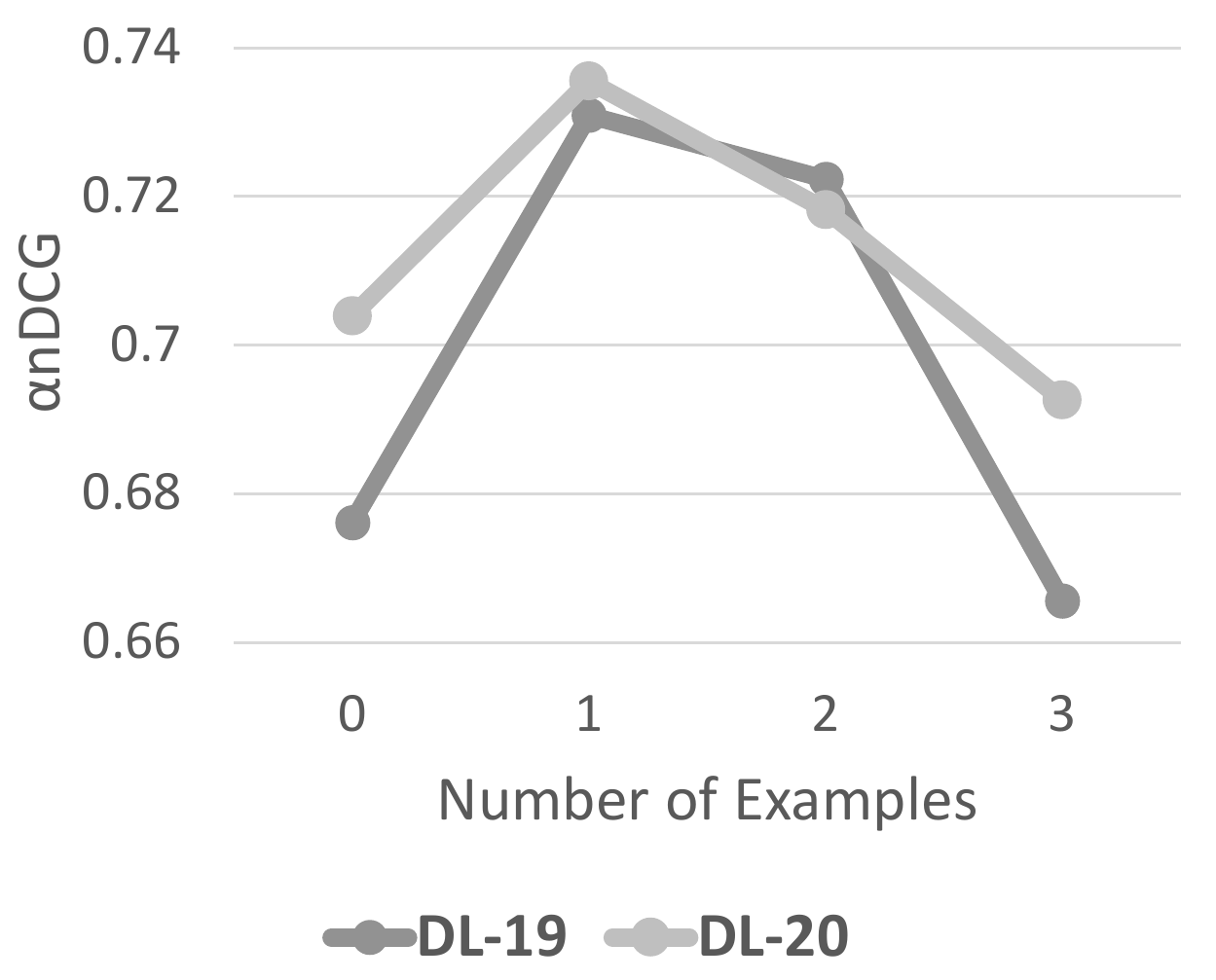}
        \caption{Analyzing the number of ICL examples for the Diversity task.}
        \label{fig:examples_diversity}
    \end{subfigure}
    \hfill %

    \begin{subfigure}[b]{0.8\columnwidth}
        \centering
        \includegraphics[width=\columnwidth]{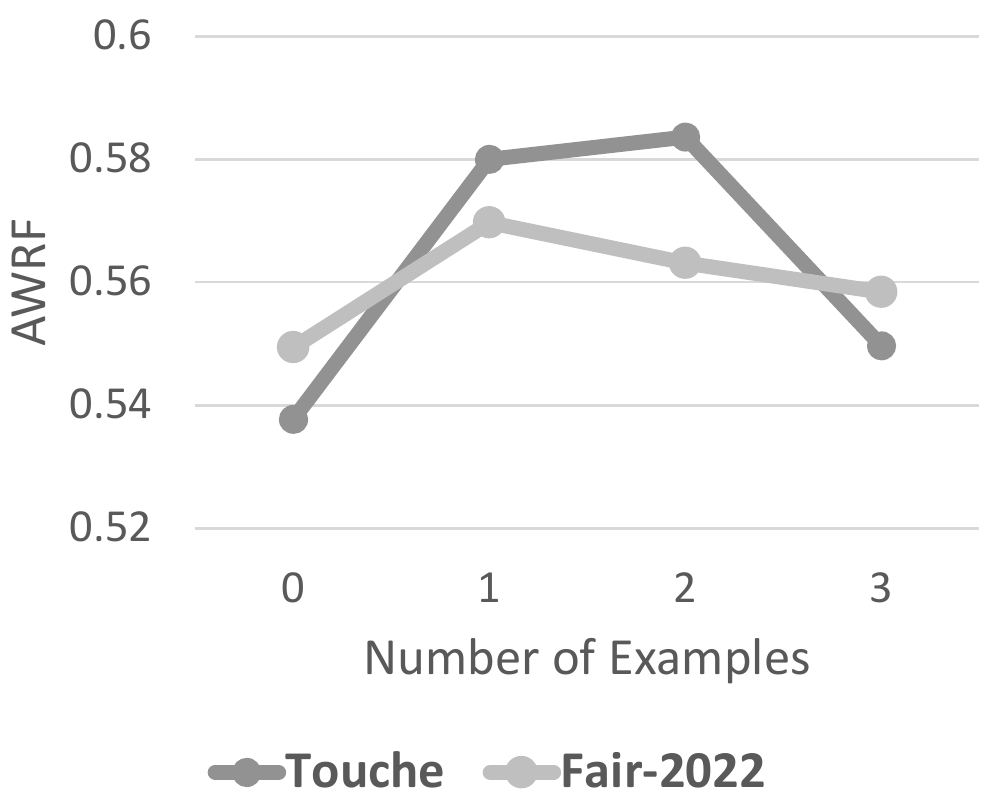}
        \caption{Analyzing the number of ICL examples for the Fairness task.}
        \label{fig:examples_fairness}
    \end{subfigure}
    
    \caption{Experimenting with the GPT-4o-mini LLM, we observe significant improvement in $\alpha$nDCG and AWRF using 1-shot in all the experiments. However, a sudden degradation in the scores after 1-shot indicates possible ambiguity in the example or LLM context length limits.}
    \label{fig:number_of_examples}
\end{figure}

We test by varying the number of examples ($k$) from 1 to 3 and found that $k=1$ yields the best performance improvement as shown in Figure \ref{fig:number_of_examples}. With each increase in the value of $k$, the context requirement doubles, since each example contains a query and its set of 20 example documents (considering a window size of 20). We observe that $k>1$ exceeds the context of small LLMs, resulting in degraded performance. For larger LLMs, we observe no significant improvement over $k=1$ performance but with increased inference time, similar to what was reported in \cite{nilanjan2024}.

\begin{figure}[t]
\centering
\begin{tcolorbox}[title={\small Sample test query from Touché dataset}]
\small
Rank the passages based on their relevance to query: 
\begin{definitions*}{Test Query from Touché}
\textbf{Should teachers get tenure?}
\end{definitions*}
\text{[1]} ... tenure is needed to $\underbrace{\text{\bbox{ protect academic freedom}}}_{\textbf{CON}}$ ...

\text{[2]} ... without tenure $\underbrace{\text{\bbox{ teachers may be fired.}}}_{\textbf{CON}}$

\text{[3]} ... it is $\underbrace{\text{\gbox{difficult to remove under-performing teachers. }}}_{\textbf{PRO}}$

Rank the 3 passages above based on their relevance to the search query. The passages should be listed in descending order using identifiers. The most relevant passages should be listed first. The output format should be [ ] > [ ], e.g., [1] > [2].
\end{tcolorbox}
\caption{The figure shows a sample input query from the Touché dataset. The ICL example of a related query from MS MARCO and its example output (balancing both relevance and pro:con parity, as shown in Figure \ref{fig:icl_fair_examples}) is used to \textit{control} the current query's reranking.}
\label{fig:icl_fair_examples_2}
\end{figure}

\subsection{LLM Configurations}
\begin{itemize}
    \item \textbf{Llama-3.1-8B-Instruct}: 8B decoder-only LLM\footnote{\href{https://huggingface.co/meta-llama/Llama-3.1-8B-Instruct}{https://huggingface.co/meta-llama/Meta-Llama-3-8B-Instruct}} with 8K context length, sufficient for in-context example ranking. We use the \texttt{``text-generation''} pipeline with the standard \texttt{bfloat16} as it is the recommended way to conduct evaluations. We use the default parameters for the rest of the experiments: \texttt{do\_sample=True},
    \texttt{temperature=0.6} and
    \texttt{top\_p=0.9}. Additionally, we set \texttt{seed=42} for all our experiments. 
    We perform these experiments locally using a single NVIDIA A100 (40GB) GPU. 
    \item \textbf{GPT-4o-mini}: Used as the primary re-ranking model\footnote{\href{https://openai.com/index/gpt-4o-mini-advancing-cost-efficient-intelligence/}{https://openai.com/index/gpt-4o-mini-advancing-cost-efficient-intelligence/}}. During inference we set the following hyperparameters: \texttt{temparature=0}, \texttt{return\_text=True} and \texttt{seed=42}. Contamination concerns are minimal since the model is not optimized for improved test scores but for behavioral modulation under auxiliary objectives.
    \item \textbf{Llama-3.1-70B-Instruct}: 70B decoder-only LLM\footnote{\href{https://deepinfra.com/meta-llama/Meta-Llama-3.1-70B-Instruct}{https://deepinfra.com/meta-llama/Meta-Llama-3.1-70B-Instruct}} with 8K context length to test behavior of targeted ICL with larger model size. We use an API service due to hardware limitations of using the full precision model locally. We use the exact same parameters as detailed in \texttt{Llama-3.1-8B-Instruct}.
\end{itemize}

 We refrain from using rank instruction-tuned models, as these models tend to exhibit greater sensitivity to prompt formatting and catastrophic forgetting of general task knowledge in ICL setups \cite{david:2024-finetune}.

\section{Effect of LLM variations} \label{appendix:llm_comparison}

\begin{table*}[t]
\centering
\small
\caption{A comparison to show the behavior of different LLMs to targeted ICL examples using similar details as shown in Table \ref{tab:nov100_results} and \ref{tab:fair100_results}. The best score across all the categories is bold, and the second-best scores are underlined.}
\label{tab:ablation_llm_compare_results}
\begin{adjustbox}{width=0.95\textwidth}
\begin{tabular}{l@{~~}lllllllllll@{~~}}
\toprule
& & \multicolumn{2}{c}{TREC DL-2019} & \multicolumn{2}{c}{TREC DL-2020} & \multicolumn{3}{c}{Touché-2020} & \multicolumn{3}{c}{Fair-2022}\\
\cmidrule(r){3-4}
\cmidrule(r){5-6}
\cmidrule(r){7-9}
\cmidrule(r){10-12}
Type & Pipeline & nDCG & $\alpha$-nDCG & nDCG & $\alpha$-nDCG & nDCG & AWRF & M1 & nDCG & AWRF & M1 \\
\midrule
\multirow{4}{*}{Baseline} & BM25 & .4795 & .4569 & .4936 & .4895 & .2530 & .4811 & .1851 & .4974 & .4901 & .2975\\
& + Llama-8B & .5977 & .5695 & .5971 &  .6357 & .2388 & .4821 & .1748 & .5658 & .4895 & .3228\\
& + Llama-70B & .7026 & .6441 & .6944 &  \textbf{.7242} & .2400 & .4994 & .1691 & .5742 & .5060 & .3278\\
& + GPT-4o-mini & .6971 & .6761 & .6826 &  .7039 & .2590 & \underline{.5377} & \underline{.1936} & .5688 & \underline{.5494} & \underline{.3428}\\
\cmidrule(r){2-12}
\multirow{3}{*}{ICL}& + Llama-8B & .6144 & .5968 & .5983 &  .6062 & .2136 & .5199 & .1486 & .5316 & .5300 & .3159\\
& + Llama-70B & .6975 & .6344 & .6906 &  .6780 & \textbf{.2625} & .4981 & .1687 & .6146 & .4910 & \underline{.3428}\\
& + GPT-4o-mini & .7124 & .7135 & .6844 &  \underline{.7228} & \underline{.2608} & \textbf{.5800} & \textbf{.2023} & .5697 & \textbf{.5697} & \textbf{.3526} \\
\midrule
\multirow{4}{*}{Baseline} & ColBERT & .7205 & .6583 & .6864 & .6385 & .2590 & .2994 & .1462 & .4854 & .2068 & .1204\\
& + Llama-8B & .7363 & .6595 & .7044 &  .6622 & .2344 & .2588 & .1169 & .5870 & .1247 & .0917\\
& + Llama-70B & \textbf{.7766} & .6788 & .7471 &  .6786 & .2347 & .2606 & .1053 & \underline{.6580} & .1850 & .1313\\
& + GPT-4o-mini & \underline{.7699} & \underline{.6850} & .7498 & .6843 & .2496 & .2197 & .1027 & .6487 & .2056 & .1466\\
\cmidrule(r){2-12}
\multirow{3}{*}{ICL}& + Llama-8B & .7116 & .6527 & .6976 &  .6443 & .2089 & .2389 & .0960 & .5183 & .2069 & .1141\\
& + Llama-70B & .7621 & .6188 & \underline{.7563} &  .6789 & .2406 & .1995 & .0909 & .6436 & .1822 & .1449\\
& + GPT-4o-mini & .7601 & \textbf{.6891} & \textbf{.7700} &  .7132 & .2508 & .2602 & .1216 & \textbf{.6606} & .2272 & .1628\\
\bottomrule
\end{tabular}
\end{adjustbox}
\end{table*}

We included test results with Llama-70B alongside Llama-8B and GPT-4o-mini to answer if and how our approach is dependent on LLM size. As observed from Table \ref{tab:ablation_llm_compare_results}, we mark that GPT-4o-mini consistently outperformed all other models when considering the target task using ICL examples, with the only exception being TREC DL-2020. Llama-8B results show that our approach works even for small models, but with limited gains both in terms of relevance and auxiliary objective. In contrast to Llama-8B, we observe that Llama-70B is strong in the relevance task; however, it does not show significant improvements in auxiliary objectives with the ICL examples. This suggests that our approach provides limited gains when the model is already superior in terms of diverse rankings. Nevertheless, under settings such as fairness, which are generally detrimental to ranking effectiveness, we can further improve and maintain nDCG. 

For the smaller re-ranker (Llama-8B), incorporating ICL examples yields an improvement in $\alpha$nDCG over the 0-shot setting on DL-19 when using BM25, though it still underperforms relative to all other baselines. When applied to a smaller model, diversity-oriented examples may inadvertently introduce less relevant or random items, negatively impacting relevance and diversity metrics. In contrast, larger models appear more capable of integrating both the diversity cues of the examples and the relevance constraints of the instructions, likely due to their greater representational capacity.

In fairness experiments, we observe a similar outcome, except for Touché under ColBERT, as seen in Table \ref{tab:fair100_results}, suggesting that even smaller LLMs are effective at modeling an auxiliary objective. The larger model, however, significantly outperforms the smaller model across the majority of settings. 

\section{Example ordering}
\label{appendix:ex_ordering}

\begin{table}[t]
\centering
\small
\caption{
Evaluating the effect of the initial ordering of example documents and ordering with the first stage over TREC DL-2019 and 2020. 
\label{tab:nov100_iclr_results}
}
\begin{adjustbox}{width=\columnwidth}
\begin{tabular}{l@{\hspace*{0.5em}}l@{~~}c@{~~}c@{}c@{~~}c@{}}
\toprule
& Example & \multicolumn{2}{c}{TREC DL-2019} & \multicolumn{2}{c}{TREC DL-2020} \\
\cmidrule(r){3-4}
\cmidrule(r){5-6}
 Pipeline & Ordering & nDCG & $\alpha$-nDCG & nDCG & $\alpha$-nDCG \\
\midrule
\multirow{2}{*}{BM25 + Diverse} & Random & .7124 & \textbf{.7135} & \textbf{.6844} &  \textbf{.7228}\\
& BM25& \textbf{.7216} & .6882 & .6823 &  .6999\\
\midrule
\multirow{2}{*}{ColBERT + Diverse} & Random & .7601 & .6891 & \textbf{.7700} &  \textbf{.7132}\\
& ColBERT & \textbf{.7784} & \textbf{.6991} & .7670 &  .7074\\
\bottomrule
\end{tabular}
\end{adjustbox}
\end{table}

\begin{table}[t]
\centering
\caption{ Measuring document order sensitivity over Touché and TREC Fair-2022, other details are similar to the evaluation shown in Table \ref{tab:nov100_iclr_results}.
\label{tab:fair100_iclr_results}
}
\begin{adjustbox}{width=\columnwidth}
\begin{tabular}{l@{\hspace*{0.5em}}l@{~~}c@{~~}c@{~~}c@{}c@{~~}c@{~~}c@{}}
\toprule
 & Example & \multicolumn{3}{c}{Touché-2020} & \multicolumn{3}{c}{Fair-2022} \\ 
\cmidrule(r){3-5}
\cmidrule(r){6-8}
Pipeline & Ordering & nDCG & AWRF & M1 & nDCG & AWRF & M1 \\ \midrule
\multirow{2}{*}{BM25 + Fair} & Random & .2608 & \textbf{.5800} & .2023 & .5697 & \textbf{.5697} & .3526 \\
 & BM25 & \textbf{.2856} & .5410 & \textbf{.2180} &  \textbf{.6029} & .5692 &  \textbf{.4013}\\
\midrule
\multirow{2}{*}{ColBERT + Fair} & Random & \textbf{.2508} & \textbf{.2602} & \textbf{.1216} & \textbf{.6606} & \textbf{.2272} & \textbf{.1628}  \\
 & ColBERT & .2444 & .2028 & .1010 &  .6554 & .2260 &  .1593\\
\bottomrule
\end{tabular}
\end{adjustbox}
\end{table}

We initially adopted random ordering for in-context examples due to its simplicity and computational efficiency. However, prior work highlights that the ordering of both examples and test documents can critically influence model behavior, introducing instability or performance degradation in tasks like ranking and generation \cite{sorensen2022-prompt,parry:2024b}. To systematically evaluate this risk, we complement random ordering with examples ordered by first-stage ranker scores (e.g., BM25 or ColBERT relevance scores). This dual approach tests whether example ordering impacts model outputs analogously to test document ordering—specifically, whether ordered examples provide clearer task conditioning while random ordering acts as a regularizer. By comparing these configurations, we isolate the effect of document ordering on the model’s ability to balance relevance and auxiliary objectives.

We observe from Tables \ref{tab:nov100_iclr_results} and \ref{tab:fair100_iclr_results}, how a randomly shuffled initial ordering of ICL examples compares to the ordering by the first stage. Intuitively, one can assume that demonstrations should closely match the exact setting in which the model is used. Our results demonstrate that stage-one ordering enhances nDCG but adversely impacts the auxiliary objective performance. 

However, we generally observe that random shuffling is robust and contributes positively to the auxiliary objectives. We attribute this to the fact that random ordering mitigates bias and enables the system to generalize effectively across diverse objectives while also enhancing the adaptability of our approach. While document order is a key factor in the robustness of supervised list-wise re-rankers~\cite{pradeep:2023}, this appears to have a reduced negative effect on exemplar-based zero-shot list-wise ranking. With likely improvements of supervised rankers in the future, these same improvements may bolster the effectiveness of in-context learning methods.

\section{Prompt Template} \label{ss:prompt-template}
\begin{figure}[t]
\centering

\begin{tcolorbox}[title={\small Prompt header}]
\small
You are RankGPT, an intelligent assistant that can rank passages based on their relevancy to the query.
\end{tcolorbox}

\begin{tcolorbox}[title={\small Example target ranking for the top-most 
similar query}]
\small
I will provide you with $\{m\}$ passages, each indicated by number identifier [ ]. As an example, the first $\{m\}$ passages are ranked based on their relevance to query: $\{Q\}$\\
\text{[1] $\{D_1\}$} \\
\text{[2] $\{D_2\}$} \\
\text{...} \\
\text{[$k$] $\{D_m\}$} \\
The $\{m\}$ passages above are ranked based on their relevance to the search query.\\
Output:
$\underbrace{\text{\gbox{[4] > [1] > ... > [6]}}}_{ \text{\textbf{Example ordering}}}$
\end{tcolorbox}

\begin{tcolorbox}[title={\small Current (Input) Query}]
\small
Rank the passages based on their relevance to query: $\{Q^c\}$\\
\text{[1] $\{D^c_1\}$} \\
\text{[2] $\{D^c_2\}$} \\
\text{...} \\
\text{[$k$] $\{D^c_m\}$} \\
Rank the $\{m\}$ passages above based on their relevance to the search query. The passages should be listed in descending order using identifiers. The most relevant passages should be listed first. The output format should be [ ] > [ ], e.g., [1] > [2].
\end{tcolorbox}

\caption{The prompt template used in our work with the header identical to that of \cite{sun2023-chatgpt}. Different from \cite{sun2023-chatgpt} our prompt allows provision to include a target ranking for a similar query. In the figure, $Q^c$ denotes the current input query, and $D^c_i$ denotes the document at position $i$ of the input ranked list, which is to be re-ranked.
}
\label{fig:icl_template}
\end{figure}

Figure \ref{fig:icl_template} shows the template for including the list-wise examples.
The sample output labeled as `Example ordering' (marked with green) in Figure \ref{fig:icl_template} refers to an ordering - a permutation map of the input - found by maximizing a given objective related to the distribution of the metadata values of each document of the input list $\langle D_1,\ldots,D_k\rangle$ retrieved for the query $Q$ which is similar to $Q_{c}$ (the current input query). This permutation of a set of input documents retrieved for a similar query is the only mechanism to `control' the output ranking for the query $Q_{c}$.

\section{Localized Queries used for ICL examples}
\begin{figure}[t]
\centering
\begin{adjustbox}{width=.95\columnwidth}
\begin{tcolorbox}[title={\small Localized Queries from TREC DL-2019}]
\small
\begin{definitions*}{Test Query}
\textbf{Do goldfish grow?}
\end{definitions*}
\begin{definitions*}{Similar Query from MS-MARCO}
\begin{enumerate}[noitemsep,leftmargin=*]
    \item How to grow goldfish faster?
    \item How big do shubunkin goldfish grow?
    \item How fast do baby goldfish grow?
    \item How big can goldfish grow?
    \item What is a goldfish?
\end{enumerate}
\end{definitions*}
\end{tcolorbox}
\end{adjustbox}

\begin{adjustbox}{width=.95\columnwidth}
\begin{tcolorbox}[title={\small Localized Queries from TREC DL-2020}]
\small
\begin{definitions*}{Test Query}
\textbf{Why is pete rose banned from hall of fame?}
\end{definitions*}
\begin{definitions*}{Similar Query from MS-MARCO}
\begin{enumerate}[noitemsep,leftmargin=*]
    \item Why is pete rose banned from mlb?
    \item Where does pete rose do autographs?
    \item How old is pete rose?
    \item When was pete rose born?
    \item How many catchers are in the hall of fame?
\end{enumerate}
\end{definitions*}
\end{tcolorbox}
\end{adjustbox}

\begin{adjustbox}{width=.95\columnwidth}
\begin{tcolorbox}[title={\small Localized Queries from Touche-2020}]
\small
\begin{definitions*}{Test Query}
\textbf{Should teachers get tenure?}
\end{definitions*}
\begin{definitions*}{Similar Query from MS-MARCO}
\begin{enumerate}[noitemsep,leftmargin=*]
    \item How many years does it take to get tenure as a teacher?
    \item What is the illinois mandate for tenured teacher evaluation?
    \item What are tenure protections?
    \item What is tenure mean?
    \item Greenspan tenure?
\end{enumerate}
\end{definitions*}
\end{tcolorbox}
\end{adjustbox}

\begin{adjustbox}{width=.95\columnwidth}
\begin{tcolorbox}[title={\small Localized Queries from Fair-2022}]
\small
\begin{definitions*}{Test Query}
\textbf{Architecture}
\end{definitions*}
\begin{definitions*}{Similar Query from MS-MARCO}
\begin{enumerate}[noitemsep,leftmargin=*]
    \item What do you do in architecture?
    \item What is it architecture?
    \item What is an architecture do?
    \item What is architectural?
    \item How roman architecture influenced modern architecture?
\end{enumerate}
\end{definitions*}
\end{tcolorbox}
\end{adjustbox}

\caption{An example showing five localized queries that are retrieved for a test query in each test collection.}
\label{fig:loc_query_examples}
\end{figure}

Using BM25, we retrieve five queries for each test query from the MS MARCO train query set. These similar queries are used as the query for ICL examples. Sample examples of such similar queries for each test are shown in Figure \ref{fig:loc_query_examples}.

\end{document}